\documentclass[superscriptaddress, aps, prl, reprint, floatfix]{revtex4-2}
\usepackage{graphicx}
\usepackage{bm}
\usepackage{xspace}
\usepackage{textcomp}
\usepackage{amsmath}
\usepackage{amssymb}
\usepackage[mathlines]{lineno}
\usepackage{gensymb}
\usepackage{xcolor}
\usepackage{enumerate}
\usepackage{mathtools}
\usepackage{siunitx}
\usepackage{booktabs}
\usepackage[normalem]{ulem}
\usepackage{}
\modulolinenumbers[3]

\renewcommand{\eqref}[1]{Eq.~(\ref{#1})}
\newcommand{\figref}[1]{Fig.\,\ref{#1}}

\newcommand{\SuppVid}[1]{\textit{SV~#1}}

\begin{document}

\title{Biomimetic Synchronization in biciliated robots}

\author{Yiming Xia}
\thanks{These authors contributed equally to this work.}
\affiliation{Beijing National Laboratory for Condensed Matter Physics and Laboratory of Soft Matter Physics, Institute of Physics, Chinese Academy of Sciences, Beijing 100190, China}
\affiliation{School of Physical Sciences, University of Chinese Academy of Sciences, Beijing 100049, China}

\author{Zixian Hu}
\thanks{These authors contributed equally to this work.}
\affiliation{Beijing National Laboratory for Condensed Matter Physics and Laboratory of Soft Matter Physics, Institute of Physics, Chinese Academy of Sciences, Beijing 100190, China}
\affiliation{School of Physical Sciences, University of Chinese Academy of Sciences, Beijing 100049, China}

\author{Da Wei}
\email{weida@iphy.ac.cn}
\affiliation{Beijing National Laboratory for Condensed Matter Physics and Laboratory of Soft Matter Physics, Institute of Physics, Chinese Academy of Sciences, Beijing 100190, China}

\author{Ke Chen}
\email{kechen@iphy.ac.cn}
\affiliation{Beijing National Laboratory for Condensed Matter Physics and Laboratory of Soft Matter Physics, Institute of Physics, Chinese Academy of Sciences, Beijing 100190, China}
\affiliation{School of Physical Sciences, University of Chinese Academy of Sciences, Beijing 100049, China}

\author{Yi Peng}
\email{pengy@iphy.ac.cn}
\affiliation{Beijing National Laboratory for Condensed Matter Physics and Laboratory of Soft Matter Physics, Institute of Physics, Chinese Academy of Sciences, Beijing 100190, China}
\affiliation{School of Physical Sciences, University of Chinese Academy of Sciences, Beijing 100049, China}

\author{Mingcheng Yang}
\email{mcyang@iphy.ac.cn}
\affiliation{Beijing National Laboratory for Condensed Matter Physics and Laboratory of Soft Matter Physics, Institute of Physics, Chinese Academy of Sciences, Beijing 100190, China}
\affiliation{School of Physical Sciences, University of Chinese Academy of Sciences, Beijing 100049, China}

\begin{abstract}  
Direct mechanical coupling is known to be critical for establishing synchronization among cilia. 
However, the actual role of the connections is still elusive - partly because controlled experiments in live samples are challenging. Here, we employ an artificial ciliary system to address this issue.
Two cilia are formed by chains of self-propelling robots and anchored to a shared base so that they are purely mechanically-coupled. The system mimics biological ciliary beating but allows fine control over the beating dynamics.
We find that the artificial cilia exhibit rich motion behaviors, depending on the mechanical coupling scheme. Particularly, their synchronous beating display two distinct modes - analogous to those observed in \textit{C. reinhardtii}, the biciliated model organism for studying synchronization. Close examination suggests that the system evolves towards the most dissipative mode. Using this guideline in both simulations and experiments, we are able to direct the system into a desired state by altering the modes' respective dissipation. Our results have significant implications in understanding the synchronization of cilia.
\end{abstract}
\maketitle

\linenumbers\relax 

Synchronization is a phenomenon across scales~\cite{Pikovsky2001}. It means that oscillators unify their
rhythm through interactions. In this way, output of individual oscillators can add up and give rise to collective behaviors on a larger scale. Ciliary motility is an archetype of such emergence. A cilium is an active eukaryotic organelle that bends periodically to pump fluid. Synchrony among thousands of beating cilia creates fluid flows on a scale ($10^{-3}$-$10^{-1}$~m) orders of magnitudes larger than a single cilium ($10^{-5}$-$10^{-4}$~m)~\cite{Wei2019,Wei2021}. Microorganisms exploit such flows to swim, and mammalians use it to transport fluid~\cite{Elgeti2015}. The efficiency of these flows are crucially affected by the exact mode of synchronization (spatial-temporal phase dynamics)~\cite{Elgeti2013}.
How cilia couple to each other to synchronize and exhibit distinct modes, is a question that has garnered decades of attention~\cite{Vilfan2006,Guirao2010,Geyer2013,Klindt2016}. In general, the coupling mechanisms fall into two categories. While hydrodynamic interaction is sufficient for some organisms~\cite{Woolley2009,Brumley2014,Pellicciota2020hydrosyncCilia}, direct mechanical connections at the ciliary bases are crucial for others, including the model organism for studying ciliary synchronization, \textit{C. reinhardtii} (CR)~\cite{Quaranta2015,Wan2016,Wei2023}. 

So far, our understanding of how mechanical connections help cilia synchronize is still limited~\cite{Soh2022}. The limitation arises from two fundamental challenges in experimenting with living samples. First, cilia operate in fluid and are closely spaced such that their hydrodynamic interactions cannot be neglected. Second, bio-mechanical coupling is difficult to isolate from the cell's ongoing physiological and bio-chemical processes for controlled experiments. An example is that the cilia of CR cells completely fail to synchronize when the cell is demembranated and reactivated \textit{in vitro}~\cite{Kamiya1987}. 


Biomimetic systems in a fluid-free environment provide a possibility to overcome these challenges. Recently, Zheng \textit{et al.} demonstrated that a chain of self-propelling robots can spontaneously oscillate and two bonded chains may even synchronize~\cite{Zheng2021}. The observed oscillation visually resembles the beating of a biological cilium and captures its overdamped nature~\cite{Dauchot2019Hexbug}. These findings suggest the biomimetic cilia to be an ideal platform for studying ciliary synchronization exclusively mediated by mechanical connections. Particularly, such systems give an opportunity to elucidate the role of ciliary connections in the emergence of distinct modes of synchrony found in ciliates~\cite{Leptos2013,Wei2023,Soh2022} and their underpinning energetics, which is of fundamental importance but largely unexplored yet.

In this Letter, we devise a biciliated robotic system with the quintessential architecture of a CR cell: two cilia anchored on the same base (body). The base is subjected to designed kinematic constraints. Meanwhile, we develop a simulation model that captures experimental observations accurately. In both experiments and simulations, our system displays both in-phase (IP) and anti-phase (AP) mode of synchronous beating that are strikingly analogous to those found in CR cells. We reveal that the emergence of and the competition between these modes are governed by the maximization of energy dissipation, \textit{i.e.} between two possible states, the system favors the one with stronger energy dissipation. 

\begin{figure}[htbp]
    \includegraphics[width=0.48\textwidth]{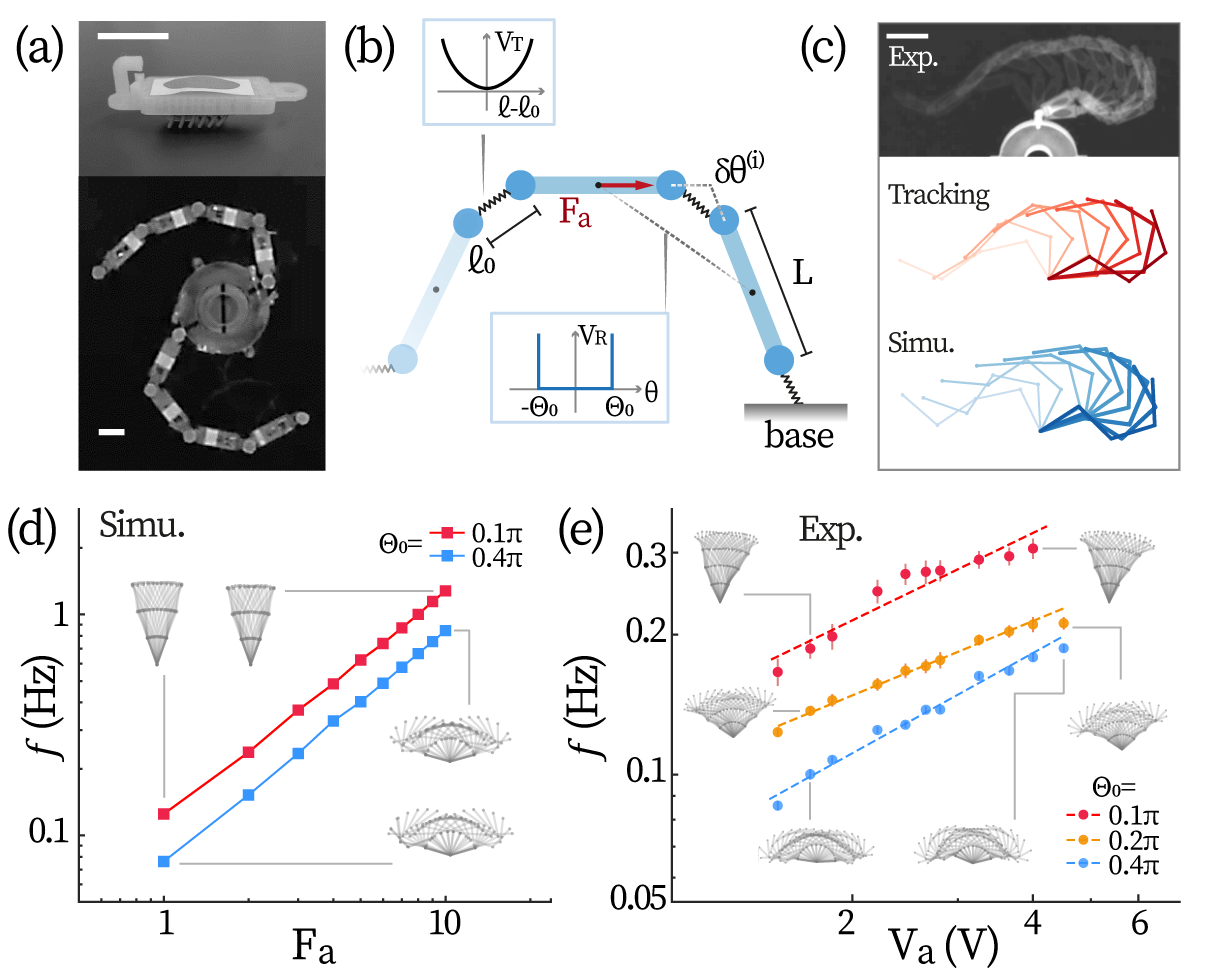}
    \vspace{-4mm}
    \caption{\textbf{Experimental and simulation systems}. (a) A Hexbug robot with 3D printed cap (upper panel) and two chains each comprising N=4 robots anchored on a common base (lower panel). Scale bars: 3~cm. (b) The simulation model. (c) Time-lapsed waveform of a single robotic cilium over half a cycle. Frequency and waveform of a single cilium vs. the self-propelling force $F_a$ ($V_a$) for different maximum bending angles $\Theta_0$, obtained by simulations (d) and experiments (e). Insets: full-cycle time-lapsed ciliary waveform under the marked conditions.} 
    \label{fig:setup}  
    \vspace{-3mm}
\end{figure}

\noindent\paragraph*{Experimental and simulation systems.} Robotic cilia are formed by connecting Hexbug (Nano\textsuperscript{\tiny\textregistered}) robots. The robot generates vertical vibration at $\sim$100~Hz which is converted into self-propulsion by its elastic forward-leaning legs (\figref{fig:setup}a top panel). An anchored chain of connected Hexbugs oscillates spontaneously due to bifurcation~\cite{Zheng2021,NoteBifurcation}, displaying waveforms analogous to biological cilia. The 3D-printed joints connecting Hexbugs are loose until neighboring units reach an angle of $\Theta_0$. Each cilium is powered with an tunable DC voltage $V_a$, and the two cilia are anchored on a shared base (\figref{fig:setup}a lower panel). In some cases, stiff metal tracks or pre-stressed elastic strings are integrated to the base to constrain its motion (\textit{e.g.} allow only translation or rotation).

Meanwhile, weights ($m$=15-500~g) are loaded on the base to tune the friction between the base and the surface of the table, which modulates the strength of mechanical coupling. Ciliary beating is recorded by videography at 46.5 fps from which each robot is tracked with custom Python scripts. From the orientations of the robots next to the base, we compute observable-independent oscillatory phases of the cilia~\cite{Kralemann2008}; and we characterize synchronization by the time fraction of phase-locked beating $\tau=t_{\rm sync}/t_{\rm total}$ ($t_{\rm total}$: the total time of observation)~\cite{Quaranta2015, Wei2023}. 

The artificial ciliary system is also studied by means of Brownian dynamics simulations, where each Hexbug is represented as an active rod (length $L$) with a constant self-propelling force $F_a$ along its long axis (\figref{fig:setup}b). Meanwhile, stochastic forces $\boldsymbol{\xi}_T$ and torques $\xi_R$ deriving from the environment with temperature $T$, are exerted on each robot but not on the base. 
Neighboring rods are connected by a short harmonic spring between their ends. The ciliary bending angle $\delta \theta$ is restricted within $[-\Theta_0,\Theta_0]$. The translational (rotational) friction coefficients for the Hexbugs and the base are respectively denoted as $\eta_0$ ($\eta_{0,R}$) and $\eta$ ($\eta_R$). Unless otherwise stated, $\eta_{(0,)R}/\eta_{(0)}=L^2/12$ is used in simulations~\cite{NoteEtaRatio}. This relation is obtained for a rod-shaped particle in the absence of hydrodynamic interaction.
The simulations reproduce the experimental results accurately (\figref{fig:setup}c). For more simulation details please see \textit{Sec.S1} in the \textit{Supplementary Materials}~\cite{SM}. 

The beating of a single cilium is characterized by its cyclic configuration (waveform), frequency ($f$), and the noise in oscillatory phase ($T_{\phi}$). In our system, the waveform is determined by the number of Hexbugs $N$, the maximal bending $\Theta_0$. Under given $N$ and $\Theta_0$, $f$ is controlled by $F_a$ ($V_a$); and $T_{\phi}$ is derived solely from $T$. Notably, in our system, control over waveform and frequency is decoupled, \textit{i.e.}, varying $F_a$ ($V_a$) does not change the waveform (\figref{fig:setup}d-e insets). This decoupling marks a key difference from the elastically connected chains in Ref.~\cite{Zheng2021}, and probably results from the inelastic connection scheme we use. The decoupling is practically advantageous for exploring the parameter space of ciliary beating. In the following, we will focus on robotic cilia formed with $N$=4 robots with $\Theta_0=0.4\pi$. 

\noindent\paragraph*{Modes of Synchronization.}
Whether two coupled oscillators can synchronize is primarily determined by the competition between their mismatch in intrinsic frequencies $\nu=f_1-f_2$ (detuning) and their coupling strength $\varepsilon$~\cite{Pikovsky2001}. Therefore, for studying how two cilia synchronize via a common anchoring base, we examine the system by scanning $\nu$ and $\varepsilon$. Practically, we fix the frequency of one cilium ($F_a=5$, $V_a$=2.4~V) and vary that of the other ($F'_a$, $V'_a$) to modulate $\nu$. As $\varepsilon$ is expected to decrease with increasing friction, assuming $\varepsilon\propto \eta^{-1}$($m^{-1}$), we scan $\eta$ or $m$ to control $\varepsilon$. Finally, in experiments, the noise is low and invariant with the driving voltage; while, in simulation, we set one of the cilia to be noise-free and the other one noisy ($T=1.16$), such that the systems display richer phase dynamics.

\begin{figure}[htbp]
    \includegraphics[width=0.48\textwidth]{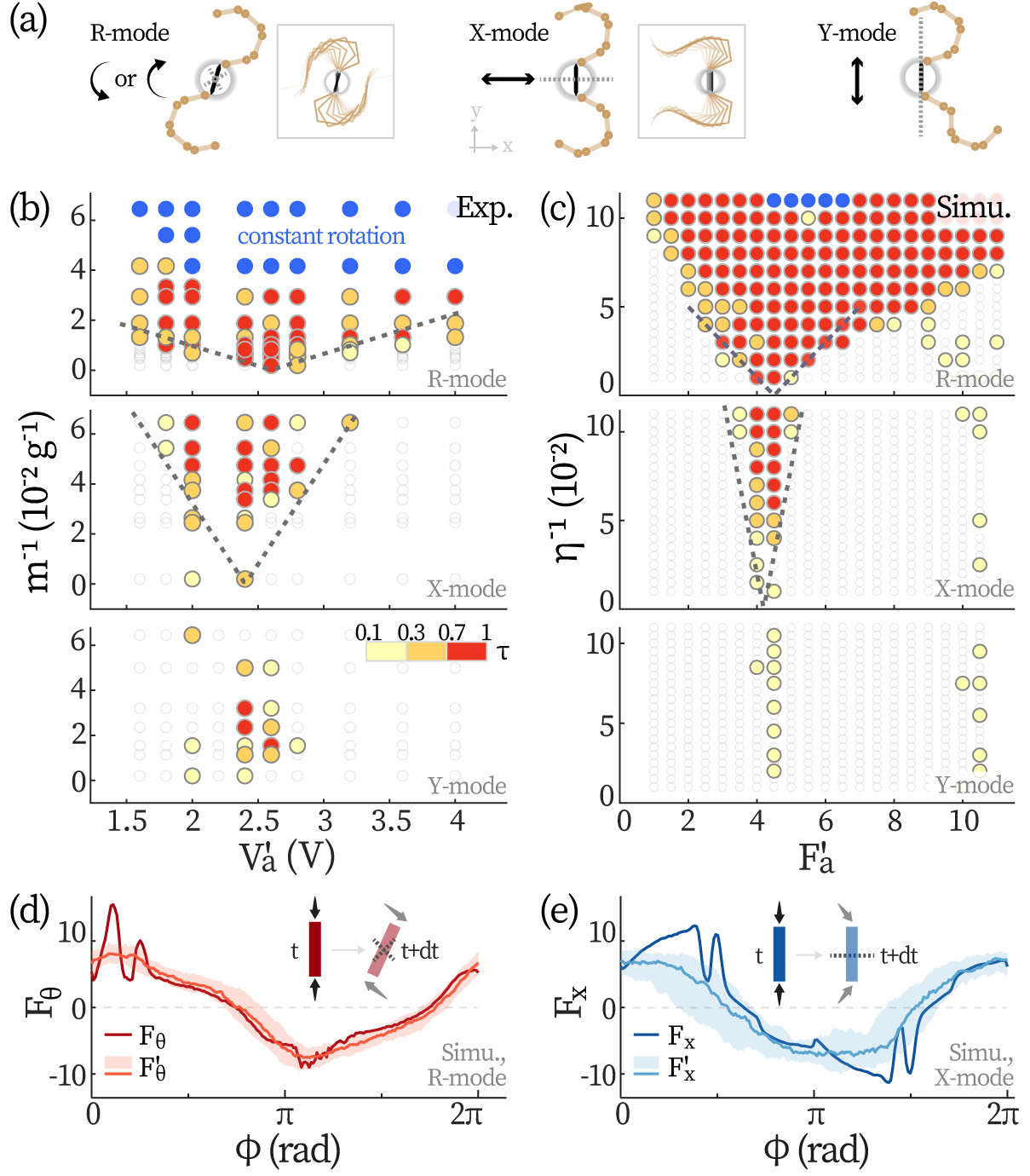}
    \vspace{-4mm}
    \caption{\textbf{Synchronization through different basal motion}. (a) Typical ciliary shapes and synchronous waveforms (boxed) supported by basal rotation (left), translation in $x$ (middle) and $y$ (right). Arnold tongue diagrams for the three basal modes obtained from experiments (b) and simulations (c). Empty points in the background are measurements where $\tau<0.1$. Dashed lines: prediction from \eqref{eq:AT} and assuming $\varepsilon\propto \eta^{-1} (m^{-1})$. Median force waveform over $\mathcal{O}$($10^2$) synchronized cycles at $\sim0$ detuning in R-mode (d) and X-mode (e). Shadings represent interquartile ranges and insets show how cilia can cooperate to maximize basal motion in each mode.}
    \label{fig:AT}
    \vspace{-5mm}
\end{figure}

We find that the basal rotation and translation promote distinct modes of synchrony (\figref{fig:AT}a). When only basal rotation is allowed (R-mode), the synchronous ciliary beating (gait) is analogous to the freestyle swimming (\textit{Supplementary Video (SV) 1}). However, when the base is confined to move along $x$-axis (X-mode), the supported gait is akin to the breaststroke of a human swimmer (\SuppVid{2}). Lastly, basal translation in $y$ (Y-mode) hardly supports any stable synchronization. Here, the $x$ axis is perpendicular to the line connecting two anchoring points of the base. The "freestyle" and "breaststroke" modes resemble the anti-phase and in-phase synchronization observed in CR~\cite{Ruffer1987,Horst1993,Leptos2013,Wei2023}, respectively.

\begin{figure}[htbp]
\includegraphics[width=0.48\textwidth]{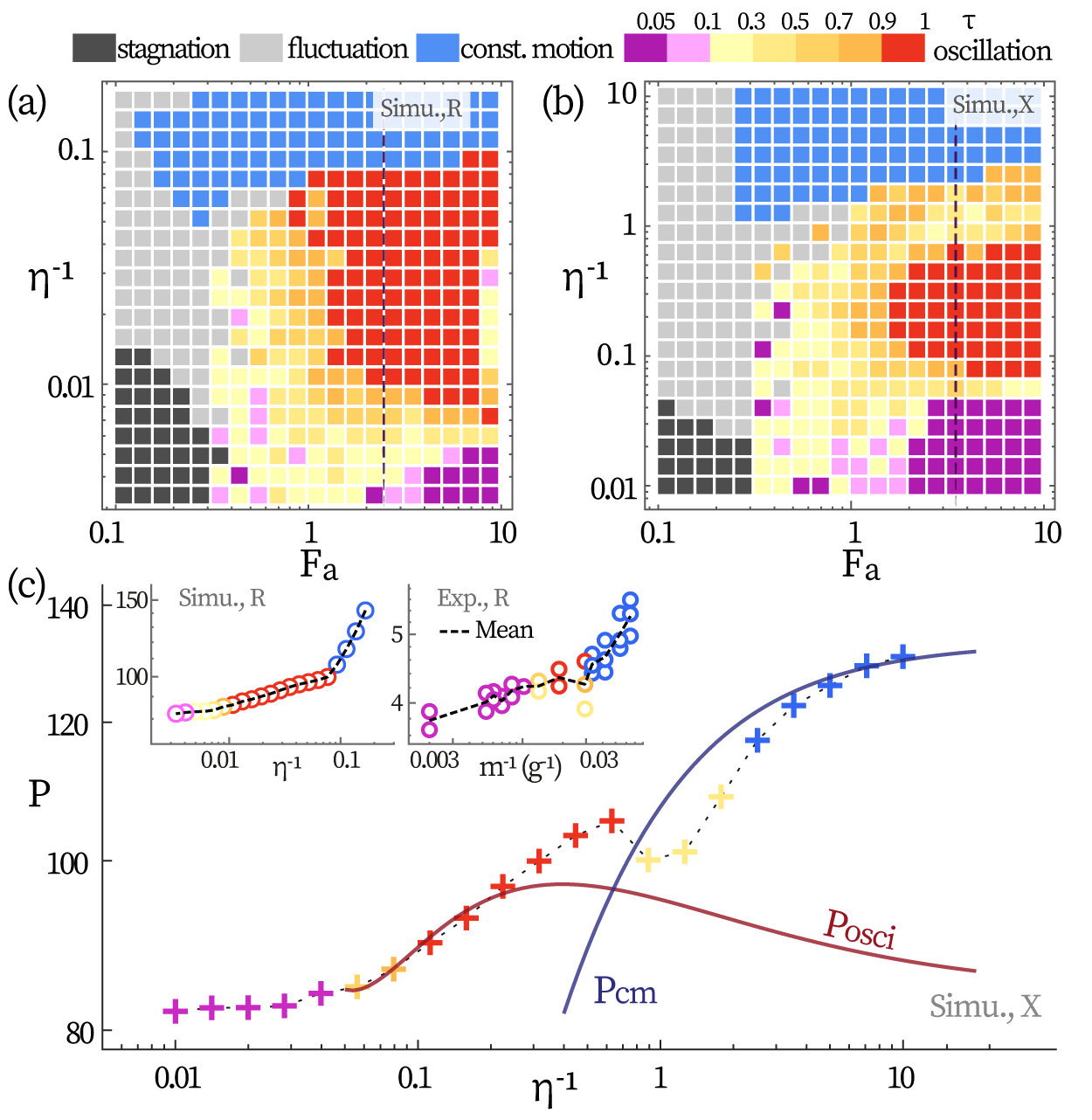}
    \vspace{-4mm}
    \caption{\textbf{Dynamic phases of two coupled cilia}. Phase diagrams of (a) the R-mode and (b) X-mode obtained from simulations. Vertical dashed lines correspond to the data shown in (c). (c) Typical dissipative power $P$ as a function of $\eta$ in X-mode ($F_a\approx3.5$). $P_{\rm cm}$ and $P_{\rm osci}$ are dissipation computed by ciliary shapes displayed in the constant motion and in the oscillation states. $P$ for R-mode is displayed in left (simulation) and right inset (experiments, $V_a\approx2.0$ V).}
    \label{fig:PD}
    \vspace{-3mm}
\end{figure}
The resultant landscape of synchronization for different basal modes are presented in \figref{fig:AT}b-c. The diagrams highlight the key feature of "Arnold tongues", that can be captured by Adler equation,
\begin{equation}\label{eq:AT}
    \dot{\Delta \phi}=2 \pi \nu - 2 \pi \varepsilon \sin \Delta\phi + \zeta(t).
\end{equation}
Here, $\Delta \phi=\phi_1-\phi_2$ denotes the oscillators' phase difference and $\zeta(t)$ is the zero-mean Gaussian white noise with $\langle \zeta (\tau + t) \zeta(\tau) \rangle = 2T_{\phi} \delta(t)$. With $\nu$ obtained from the frequency-driving relation in \figref{fig:setup}d-e and the condition $|\varepsilon| \geq |\nu|$ for the emergence of synchrony, Eq.~(\ref{eq:AT}) well reproduces the region of synchronization for the R-mode and X-mode (dashed lines in \figref{fig:AT}b-c). Here, $\varepsilon$ is used as a free parameter, yielding $\varepsilon_X=0.1\eta^{-1}$ for the X-mode and $\varepsilon_R\approx5\varepsilon_X$ for the R-mode. As the Y-mode cannot support stable synchronization, it will not be discussed. Moreover, \eqref{eq:AT} also captures the steady-state phase difference $\delta\phi=\sin^{-1}(\nu/\varepsilon)$ between synchronized cilia in simulations (\textit{Sec.S2} in ~\cite{SM}).

Although the "freestyle" (in R-mode) and the "breaststroke" (in X-mode) gaits bear distinct appearances, their driving forces exerted by the cilia on the base (\textit{i.e.}, the azimuthal component $F_\theta$ for the R-mode and the $x$ component $F_x$ for the X-mode) actually experience the same in-phase coordination, see \figref{fig:AT}d-e. The interplay between the ciliary forces qualitatively explain why $\varepsilon_R > \varepsilon_X$ as follows. In this system, one cilium exerts force on the base and generates basal motion that influences the beating of the cilium on the other end. In this way, the two cilia interact. Naturally, the larger the transmitted motion is, the stronger the coupling will be. In the X-mode, a given force $F$ applied at one end generates, per unit time, a displacement of $F/\eta$. However, in the R-mode, the resultant displacement at the other end is much larger ($3F/\eta$), thus consistent with $\varepsilon_R > \varepsilon_X$.

Besides the synchronous beating, when friction of the base $\eta$ is extremely low, the system in the R-mode evolves into in a state of constant rotation (blue bullets in \figref{fig:AT}b-c). In this case, the system persistently rotates in a direction at a uniform speed, with the cilia maintaining a stable configuration. In order to explore other possible dynamic states, we set the cilia to be 10\%-detuned ($F_a'=1.1F_a$) and equally noisy ($T=0.58$), and scan $F_a$ and $\eta$ over three orders of magnitude in simulations. The system's complete state diagrams in the R- and X-mode are respectively displayed in \figref{fig:PD}a-b and they resemble each other qualitatively. The diagrams' left sides are fluctuation regimes where the noise $T$ dominates over $F_a$. Here, cilia either wiggle without a well-defined frequency (gray) or simply stagnate (dark gray). When $F_a$ overcomes the noise, as $\eta$ decreases, two cilia evolve from beating independently (dark purple) to beating synchronously (yellow to red). Eventually, for sufficiently low $\eta$, the R-mode and X-mode are respectively dominated by constant rotation and constant translation (blue). Please see \SuppVid{3-5} for the mentioned states and \textit{Sec.S3} in~\cite{SM} for how they are quantitatively labeled.

\begin{figure*}[htbp]
    \includegraphics[width=1\textwidth]{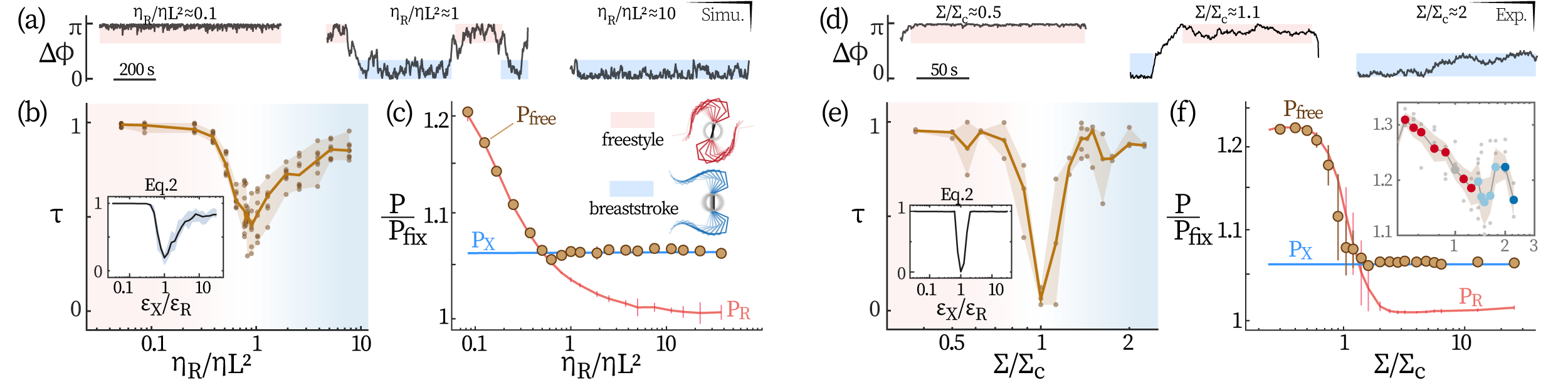}
    \vspace{-4mm}
    \caption{\textbf{Competition between the modes of synchrony and the underlying energetics}. (a,d) Ciliary phase difference $\Delta\phi$ under varying conditions, folded to [0,$\pi$] range. The freestyle ($\Delta\phi\approx\pi$) and breaststroke gait ($\Delta\phi\approx0$) are marked respectively red and blue. (b,e) The systems' total time fraction of synchrony $\tau$ with corresponding theoretical predictions by \eqref{eq:IPAP}. Background colors mark the dominant gait; dots are single measurements whose medians and interquartile ranges are respectively shown as lines and brown shadings. (c,f) The systems' total dissipative power when the base is free (circles), or geometrically constrained (R- and X-mode). Note that the data in (f) are the simulation results while that in that inset are experiments. The symbols in inset are colored by the gait dominance. $P_{\rm fix}$: dissipation on a fixed base; $\Sigma_c$: critical stress where gait dominance changes; error bars in (c) and (f) are 1 std. over N=5 and 13 datasets, respectively.}
    \label{fig:IPAP}
    \vspace{-3mm}
\end{figure*}

\paragraph*{Energetics.}
Such rich state diagrams naturally raise a question: energetically, what determines the stability of these states and their inter-transitions?
Inspired by previous studies~\cite{Klindt2016,Izumida2016} that revealed a positive association between oscillators' coupling strength and energy dissipation, we examine how dissipation is involved in the system's evolution. We find that, between two possible states, the system will evolve into the more dissipative one. In the overdamped system, heat dissipation is related to the entropy production of the medium, which corresponds to total entropy production in the stationary state~\cite{Seifert2005,Seifert2012}. In this light, our finding conforms to the maximum entropy production principle found in some other nonequilibrium dissipative systems~\cite{Martyushev2006Review}.

The system's dissipation is computed as $P=\langle\sum_i (\bm{F}_i \cdot \bm{v}_i + M_i\omega_i)\rangle$~\cite{Seifert2005}. The subscript $i$ ranges over all robot units and the base; $\bm{v}$ and $\omega$ denote velocity and angular velocity respectively. For details see \textit{Sec.S4} in \cite{SM}.
Representative traces of $P$ underlying the transition from oscillation to constant motion are displayed in \figref{fig:PD}c (corresponding to the vertical dashed lines in \figref{fig:PD}b). Overlaid on the system's actual dissipation $P$ are the estimated dissipation of ciliary oscillations $P_{\rm osci}$ and that of constant motion $P_{\rm cm}$ (\textit{Sec.S5} in~\cite{SM}). Clearly, when constant motion becomes more dissipative ($P_{\rm cm}>P_{\rm osci}$), it replaces synchronous oscillation as the dominant state. In R-mode, the picture is qualitatively the same and we can compare simulation data (left inset) with experimental results (right inset \figref{fig:PD}c). The experimental evolving trend of dissipation is accurately captured by simulations.

The results so far show that, between two possible states, the system favors the one with stronger dissipation. In this light, beating cilia coupled through a freely-moving base, which supports breaststroke and freestyle synchrony at the same time, could be directed into either mode by tuning the modes' relative dissipation. We now test this scenario.

In simulations, we fix $\eta$ and hence keep the dissipation of breaststroke gait ($P_X$) constant. Meanwhile, $\eta_R$, the friction coefficient for basal rotation, is decoupled from $\eta$ and varied to modulate dissipation of freestyle gait ($P_R$). Under given $\eta_R$, $P_R$ and $P_X$ is measured by putting the system in R-mode and X-mode respectively. Then, we free the base from any hard constraints, \textit{i.e.}, it can now simultaneously rotate and translate. The system's actual dissipation with the free base is measured as $P_{\rm free}$. 

From left to right, \figref{fig:IPAP}a shows the breaststroke (blue) overtaking the freestyle (red) as the system's dominant gait under increasing $\eta_R$ (nondimensionalized by $\eta L^2$). Near the transition point, ciliary synchronization becomes less stable - as marked by the drop in total synchronized time fraction $\tau$ (\figref{fig:IPAP}b). Since $\eta_R$ does not vary detuning, the destabilization must result from a decreased coupling strength, \textit{i.e.}, the presence of two possible gaits weakens the total coupling. This effect is captured by including the in-phase (IP) and anti-phase (AP) coupling simultaneously into \eqref{eq:AT}: 
\begin{equation}
    \dot{\Delta \phi}=2 \pi \nu - 2 \pi \varepsilon_X\sin \Delta\phi - 2\pi\varepsilon_R \sin (\Delta\phi-\pi) + \zeta(t),
    \label{eq:IPAP}
\end{equation}
with $\varepsilon_{X}$ and $\varepsilon_{R}$ (both$>$0) the IP and AP coupling strengths, respectively. In this way, a free base provides an effective coupling of $\varepsilon_X-\varepsilon_R$, \textit{i.e.}, synchrony emerges when $|\varepsilon_X-\varepsilon_R|>|\nu|$. The equation reproduces the evolving trend of $\tau$ ($\varepsilon_X=2.5\nu$, $T_{\phi} = 0.9$), see \figref{fig:IPAP}b inset. Finally, we benchmark the system's actual dissipation $P_{\rm free}$ against those under constraints, $P_R$ and $P_X$ (\figref{fig:IPAP}c). Clearly, maximal dissipation is the underlying basis of gait competition. 

Experimentally, it is challenging to decouple $\eta_R$ from $\eta$ and we thus employ another approach to modulate $P_R$. An elastic string is integrated into the base to limit its rotation (\SuppVid{6}). The string is pre-stressed such that its recovery force ($\Sigma$) is near-constant when the base rotates (as the rotation-induced stretch is negligible compared
to the pre-stretch). Similar to increasing $\eta_R$, raising $\Sigma$ also makes breaststroke the dominant gait (\figref{fig:IPAP}d). We denote the transitional stress as $\Sigma_c$ and scale $\Sigma$ with it. Near the transition, $\tau$ drops more sharply than in \figref{fig:IPAP}b, which attributes to a lower noise in experiments ($T_{\phi}=0.06$), see \figref{fig:IPAP}e and inset.
In simulations, systems with constant recovery forces also favors the more dissipative mode, see \figref{fig:IPAP}f main figure. The corresponding experimental data is displayed in the inset, aligning with the simulation trend. 

\paragraph*{Discussion.}
We have clarified the role of mechanical coupling in ciliary synchronization and the results have important implications for understanding the phase dynamics of biological biciliates. For example, while the wildtype CR only beats stably in breaststroke, a well-known CR mutant (\textit{ptx1}) can switch randomly and frequently between breaststroke and freestyle gait. How such bistability emerges in this mutant has been a pending question for decades~\cite{Horst1993,Liu2018,Guo2021,Wei2023}, and our findings indicate that it necessitates two ingredients: both gaits being similarly dissipative and the system being sufficiently noisy. Furthermore, the realistic mechanical couplings in cells involve complex networks of intra-cellular fibers~\cite{Quaranta2015,Wan2016,Soh2022}, whose effective coupling strengths and related energetics deserve attention. Our setup can be easily customized to include and test more complex coupling schemes, promoting future studies.

We thank Xiangjun Xing for helpful discussions. This work was supported by the National Natural Science Foundation of China (No. T2325027, 12274448, 12174434, 11874395, 12074406, T2221001, 12204525), National Key R\&D Program of China (2022YFF0503504),  the Strategic Priority Research Program of Chinese Academy of Sciences (No. XDB33000000), and the China Postdoctoral Science Foundation (No. YJ20200202).



\bibliographystyle{apsrev4-2}
\bibliography{./reference.bib}

\begin{thebibliography}{32}
\bibitem [{Not({\natexlab{a}})}]{NoteBifurcation}%
  \BibitemOpen
  \bibinfo {note} {Hopf-bifurcation is shown to underpin the emergence of
  spontaneous oscillation, which requires $F_a/C$ to exceed a threshold, with
  $C$ the stiffness of inter-Hexbug connection. Our setup represents the
  extreme case with zero-elasticity ($C=0$) such that there is no threshold for
  activity ($F_a$).}\BibitemShut {Stop}%
\bibitem [{Not({\natexlab{b}})}]{NoteEtaRatio}%
  \BibitemOpen
  \bibinfo {note} {The frictional power for a rod-shaped object (length L)
  rotating at angular speed $\omega$ is $\eta L^2\omega^2/12$. Writing the
  power in the form of $\eta_R \omega^2$, one obtains
  $\eta_R/\eta=L^2/12$.}\BibitemShut {Stop}%
\end{thebibliography}
\end{document}